\documentclass[conference]{IEEEtran}
\IEEEoverridecommandlockouts
\usepackage{cite}
\usepackage{url}
\usepackage{amsmath,amssymb,amsfonts}
\usepackage{algorithmic}
\usepackage{graphicx}
\usepackage{textcomp}
\usepackage{xcolor}
\def\BibTeX{{\rm B\kern-.05em{\sc i\kern-.025em b}\kern-.08em
    T\kern-.1667em\lower.7ex\hbox{E}\kern-.125emX}}
\begin{document}

\title{SWE-Arena: An Interactive Platform for Evaluating Foundation Models in Software Engineering}

\author{\IEEEauthorblockN{Zhimin Zhao}
\IEEEauthorblockA{\textit{School of Computing} \\
\textit{Queen's University}\\
Kingston, Canada \\
z.zhao@queensu.ca}
}

\maketitle

\begin{abstract}
Foundation models (FMs), particularly large language models (LLMs), have shown significant promise in various software engineering (SE) tasks, including code generation, debugging, and requirement refinement. Despite these advances, existing evaluation frameworks are insufficient for assessing model performance in iterative, context-rich workflows characteristic of SE activities. To address this limitation, we introduce \emph{SWE-Arena}, an interactive platform designed to evaluate FMs in SE tasks. SWE-Arena provides a transparent, open-source leaderboard, supports multi-round conversational workflows, and enables end-to-end model comparisons. The platform introduces novel metrics, including \emph{model consistency score} that measures the consistency of model outputs through self-play matches, and \emph{conversation efficiency index} that evaluates model performance while accounting for the number of interaction rounds required to reach conclusions. Moreover, SWE-Arena incorporates a new feature called \emph{RepoChat}, which automatically injects repository-related context (e.g., issues, commits, pull requests) into the conversation, further aligning evaluations with real-world development processes. This paper outlines the design and capabilities of SWE-Arena, emphasizing its potential to advance the evaluation and practical application of FMs in software engineering.
\end{abstract}


\section{Introduction}

Foundation models (FMs), such as large language models (LLMs), have significantly advanced software engineering (SE) tasks, including code generation~\cite{vaithilingam2022expectation}, debugging~\cite{chenteaching}, and requirement refinement~\cite{borg2024requirements}. However, their effectiveness often depends on complex, iterative workflows, which typically require models to process user feedback, revise responses dynamically, and maintain contextual coherence over multiple turns. Such context-heavy, real-world SE scenarios challenge traditional evaluation methods. Existing frameworks, such as Chatbot Arena\footnote{\url{https://lmarena.ai/?leaderboard}\label{footnote:Chatbot Arena}}, WebDev Arena\footnote{\url{https://web.lmarena.ai}\label{footnote:WebDev Arena}} and Copilot Arena\footnote{\url{https://github.com/lmarena/copilot-arena}\label{footnote:Copilot Arena}}, have introduced pairwise comparisons to evaluate model preferences. While these frameworks offer utility, they fall short of meeting the iterative and context-specific demands of SE evaluations.

For example, Chatbot Arena\footref{footnote:Chatbot Arena}, while widely referenced, suffers from several limitations that restrict its utility in SE domain. Firstly, it is not open-sourced, which reduces transparency and limits opportunities for community-driven innovation. Moreover, its exclusive reliance on Elo score~\cite{albers2001elo} and average win rate (the probability of a model winning) provides a narrow perspective in model comparison. Additionally, Chatbot Arena does not support multi-round interactions, a critical feature for evaluating FMs in real-world SE tasks. WebDev Arena\footref{footnote:WebDev Arena} and Copilot Arena\footref{footnote:Copilot Arena} also faces similar challenges. 

To address these limitations, we propose \emph{SWE-Arena}\footnote{\url{https://huggingface.co/spaces/SE-Arena/Software-Engineering-Arena}}, an interactive platform tailored for the evaluation of FMs in SE tasks. SWE-Arena enhances existing frameworks by incorporating multi-round dialogues and a transparent evaluation methodology, allowing engineers to benchmark FMs across diverse SE workflows. In addition, SWE-Arena introduces a new feature called \emph{RepoChat} that automatically extracts and injects repository-level context into the conversation, enabling deeper and more realistic benchmarking of FMs for day-to-day software engineering scenarios. This approach fills critical gaps in current evaluation practices, advancing the field toward more nuanced and context-aware assessments. To our knowledge, SWE-Arena is the first platform to integrate these capabilities for SE-specific evaluations. This paper introduces SWE-Arena, outlines its key features (including RepoChat), and highlights its potential to address the unique challenges of FM evaluation in SE.

\section{Background}

\subsection{Static Benchmarks for Model Evaluation}

The majority of existing benchmarks for evaluating machine learning (ML) models rely on static, ground truth-based datasets, typically featuring multiple-choice questions or predefined test cases. While these benchmarks cover diverse topics such as language understanding, coding, and logical reasoning, they often fail to capture real-world SE workflows. Notable SE benchmarks include BigCodeBench~\cite{zhuo2024bigcodebench}, DevOps-Eval\footnote{\url{https://github.com/codefuse-ai/codefuse-devops-eval}}, EvalPlus~\cite{liu2024your}, Long Code Arena\footnote{\url{https://huggingface.co/spaces/JetBrains-Research/long-code-arena}}, and SWE-bench~\cite{jimenez2023swe}. Although many of these benchmarks are open-source, they often rely on predefined datasets rather than user-driven evaluations, limiting adaptability to evolving SE tasks. SWE-Arena addresses this gap by supporting user-generated real-time evaluations in diverse SE scenarios. Static benchmarks are valuable for controlled evaluations but fall short in scenarios requiring interactive and iterative problem-solving~\cite{chiang2024chatbot}. Recognizing this limitation, we introduce SWE-Arena, the first open, large-scale crowd-sourced platform designed to evaluate FMs through live human interaction in SE tasks. SWE-Arena leverages dynamic evaluations to address the gaps left by static approaches, enabling more realistic assessments of model performance in real-world SE contexts.

\subsection{Pairwise Comparisons for Model Evaluation}

Researchers establish ML model evaluation based on human preferences as a well-established practice~\cite{bai2022training,christiano2017deep}. Pairwise comparisons mitigate subjective bias by asking users to choose between two options rather than assigning absolute scores~\cite{zheng2023judging}. Methods such as the Bradley-Terry model~\cite{hunter2004mm} and the Elo rating system~\cite{albers2001elo} aggregate pairwise outcomes into meaningful rankings. Evaluation platforms, such as Chatbot Arena\footref{footnote:Chatbot Arena}, implement pairwise comparison frameworks, allowing users to engage in direct battles between anonymous models. 

However, Chatbot Arena's current implementation primarily presents the average win rate alongside the Elo score, offering a limited perspective on model evaluation. To address these limitations, SWE-Arena expands the scope of pairwise comparisons for FMs in SE tasks by incorporating a broader range of evaluation metrics. In addition to traditional metrics such as the Elo score, average win rate, and Bradley-Terry coefficients, SWE-Arena integrates advanced metrics, including eigenvector centrality value~\cite{bonacich1987power}, PageRank score~\cite{page1999pagerank}, and Newman modularity score~\cite{newman2004fast}, to provide a more robust and multidimensional comparison framework.

Furthermore, SWE-Arena introduces two novel metrics:
\paragraph{Model Consistency Score (MCS)} that quantifies the percentage of self-play battles where the same model produces outputs of the same quality for identical inputs:
$$MCS = \frac{D}{N} \times 100\%$$
Here, $D$ represent draws against itself, while $N$ is the total number of self-play matches. Since draws in self-play can only occur when the model gives answers with similar quality to the same question, this formula elegantly captures the frequency of consistent model outputs.
\paragraph{Conversation Efficiency Index (CEI)} which evaluates models based on both the outcome of comparisons and the number of interaction rounds required to reach a decision:
$$\text{CEI} = \frac{\sum_{i=1}^{n}\frac{s_i}{n_i}}{\sum_{i=1}^{n}\frac{1}{n_i}}$$
Where $n_i$ represents the number of chat rounds for a single conversation, $s_i$ represents the outcome score assigned in a single user vote, with
$$
s_i = 
\begin{cases}
1, & \text{for win} \\
0.3, & \text{for draw (both working well)} \\
-0.3, & \text{for draw (both not working)} \\
-1, & \text{for loss}
\end{cases}
$$
This metric inherently discounts performance based on conversation length. Models requiring fewer rounds to achieve positive outcomes score higher than those that need extensive interaction. The nonzero value for the draws ($0.1$) ensures that the effort expended to achieve even a neutral result is adequately accounted for in the efficiency calculation, as using zero would make the round count irrelevant for the comparisons drawn.

Using these advanced metrics, SWE-Arena provides a richer and more nuanced understanding of model performance. This multidimensional evaluation framework empowers users to make informed decisions, balancing global dominance (captured by metrics such as eigenvector centrality and PageRank) with domain-specific competencies (revealed by Newman modularity), consistency characteristics (quantified by the model consistency score) and conversation efficiency (measured by the conversation efficiency index).

\subsection{Iterative and Multidimensional Characteristics of Software Engineering Tasks}

Software engineering is a multifaceted discipline that extends beyond code generation. It includes activities such as requirements engineering, release engineering, software project management, and collaborative design~\cite{biolchini2005systematic}. These diverse domains require evaluation platforms capable of addressing a wide variety of tasks. However, existing frameworks, such as Chatbot Arena\footref{footnote:Chatbot Arena}, WebDev Arena\footref{footnote:WebDev Arena}, and Copilot Arena\footref{footnote:Copilot Arena}, primarily focus on specific aspects, such as code generation and completion. While valuable, these platforms do not capture the full breadth of SE tasks, leaving a gap in understanding the broader scope and challenges of SE workflows.

A crucial factor that distinguishes SE tasks from other domains is their inherently iterative and dynamic nature. For example, in an interactive debugging session, a model may initially suggest a fix, but user feedback might indicate an unexpected failure, requiring the model to refine its solution over multiple exchanges. Similarly, in requirement engineering, an initial model-generated specification may undergo multiple refinements based on stakeholder feedback before reaching an acceptable state. These scenarios highlight the need to evaluate FMs on their ability to sustain meaningful multi-round interactions.

By addressing both the breadth of SE domains and the iterative nature of SE workflows, SWE-Arena bridges critical gaps in existing evaluation methodologies. Its support for multi-round interactions and a wide range of SE tasks—augmented by repository-level contextual data through RepoChat—provides a more comprehensive and actionable framework for benchmarking foundation models in SE.

\section{Interface}

SWE-Arena collects user feedback through crowd-sourced evaluations to rank FMs in SE tasks. Our goal is to provide an intuitive and user-friendly interface to minimize friction for users who contribute data. The platform adopts a pairwise comparison mechanism in which users compare the responses of two models and vote for the better one, without requiring absolute scores. This approach ensures consistency among the various participants and simplifies the evaluation process.

\subsection{First Round: Optional Repository-Aware Interaction}

\begin{figure}
\centering
\includegraphics[width=\linewidth]{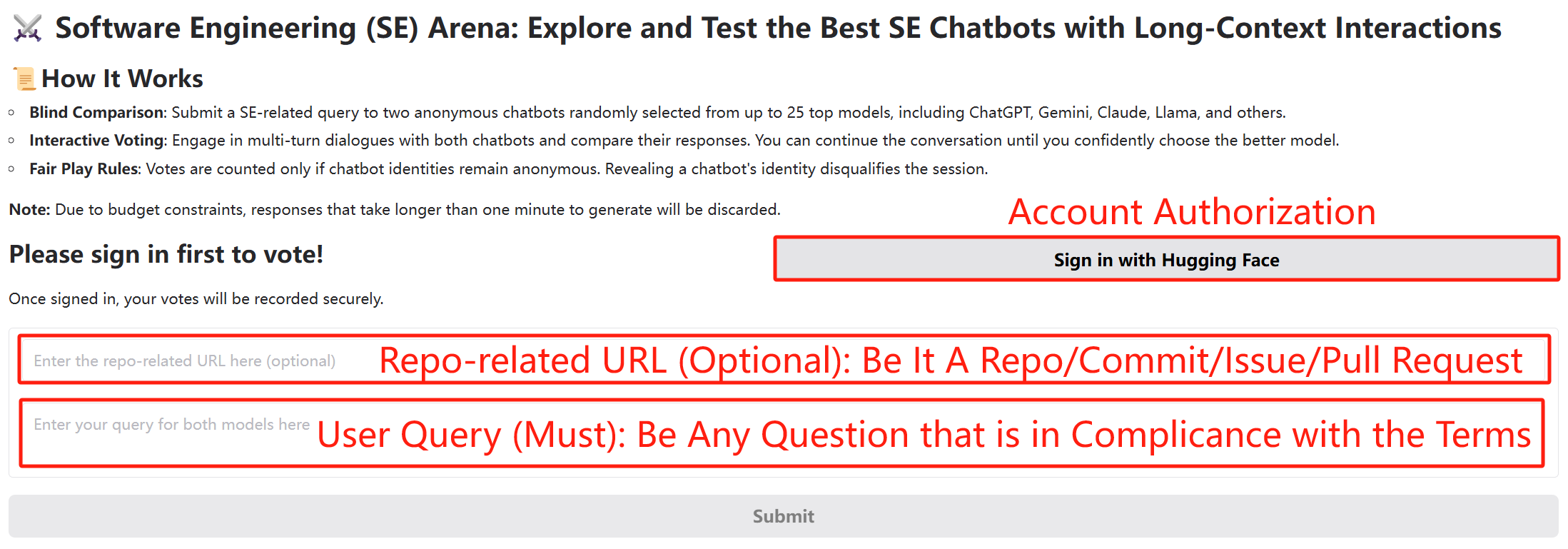}
\caption{SWE-Arena interface for the first-round user query.}
\label{fig:arena-first}
\centering
\end{figure}

Figure~\ref{fig:arena-first} illustrates the components of the SWE-Arena panel during the first round of interaction. Users begin by signing in, after which they can enter their first prompt in the designated text box. The prompts typically relate to SE issues, such as debugging, re-finement of requirements, or code review. Once submitted, two anonymous FMs are randomly selected from a preconfigured model pool to ensure fair pairwise comparisons.

To enhance the applicability of the evaluation, SWE-Arena introduces RepoChat, an optional feature that injects repository-related context into the conversation. Users can provide a repository-related URL (e.g., a GitHub or GitLab repository, issue, discussion, commit, or pull/merge request). If supplied, SWE-Arena retrieves relevant metadata, such as repository description, programming language, issue discussions, or commit diffs. These contextual data are added to the user query, forming a consolidated prompt. If no repository URL is provided, SWE-Arena functions as usual, processing only user direct input.

RepoChat reduces the need for users to manually provide extensive contextual details, ensuring that model evaluations reflect real-world SE workflows more accurately. By seamlessly integrating repository-aware context, SWE-Arena extends beyond traditional static prompts, enabling more meaningful assessments of FMs in software engineering tasks.

\subsection{Multi-Round Conversation Panel}

\begin{figure}
\centering
\includegraphics[width=\linewidth]{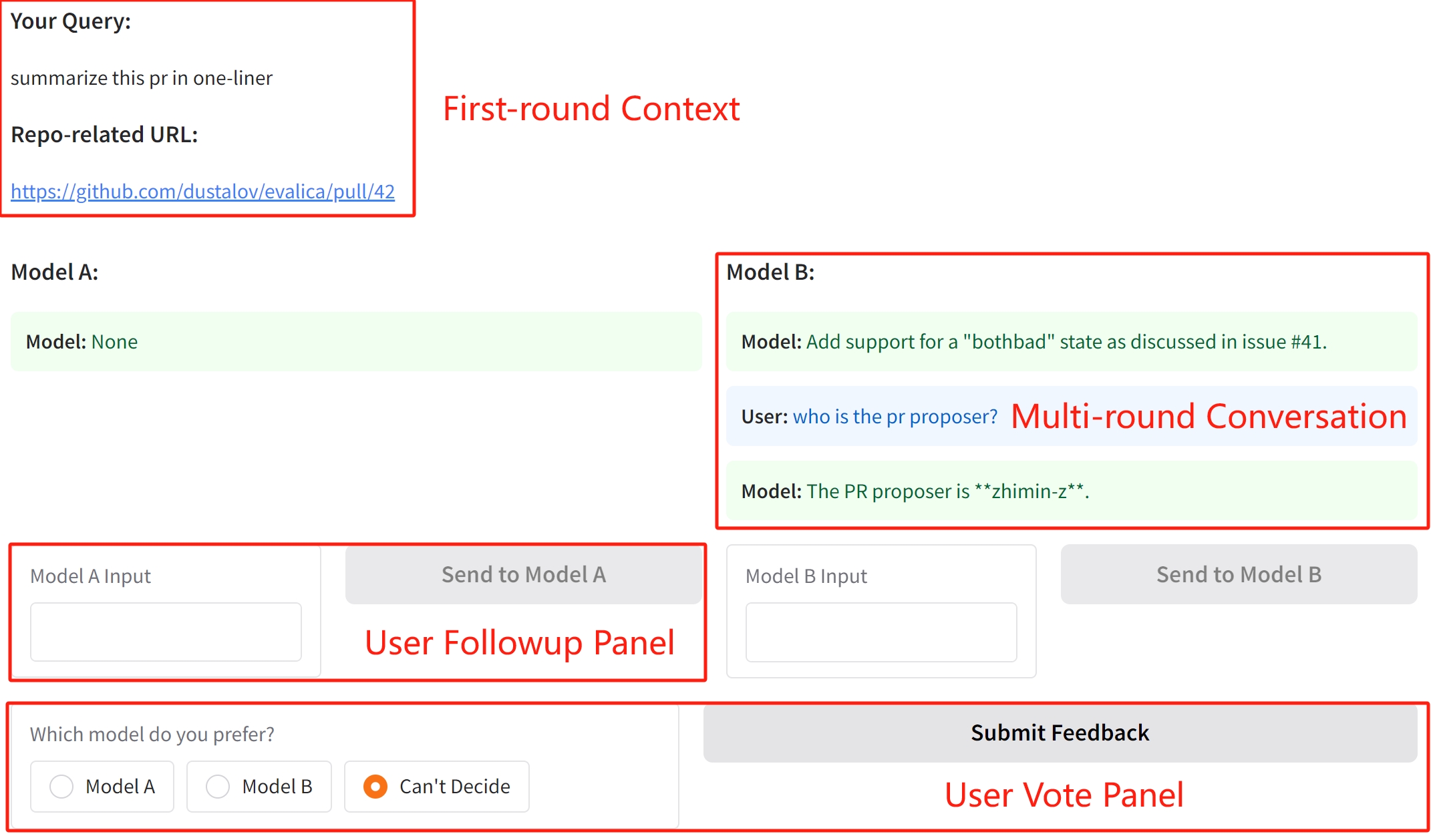}
\caption{SWE-Arena interface for multi-round conversations between users and FMs.}
\label{fig:arena-multi}
\centering
\end{figure}

After the initial round, the interface transitions into a multi-round conversation panel, as shown in Figure~\ref{fig:arena-multi}. This feature allows users to continue the dialogue with both models by asking follow-up questions based on their initial responses. The iterative process enables a deeper evaluation of each model's ability to handle context-dependent and long-term SE workflows. Users can submit their votes at any time. To mitigate potential bias due to initial impressions (e.g., favoring the first-turn response), SWE-Arena introduces a re-assessment feature, allowing users to modify their votes after reviewing multiple turns.

All interactions remain anonymous, with chatbot identities concealed throughout the session. Votes are deemed valid only if the submitted queries are relevant to SE tasks. To enforce this, SWE-Arena implements a guardrail mechanism that uses \texttt{gpt-5-nano}\footnote{\url{https://platform.openai.com/docs/models/gpt-5}} to automatically filter out non-SE-related prompts at submission time, ensuring that evaluations remain focused on meaningful SE scenarios.

When user input exceeds the maximum context window supported by the tested models, SWE-Arena applies a first-in, first-out (FIFO) strategy to remove the oldest interactions, ensuring the conversation remains within the allowable context window. For practicality, the maximum response time for a model is capped at one minute. If a model exceeds this limit, the user cannot submit additional input for that model.

\subsection{Leaderboard Integration and Data Privacy}

SWE-Arena provides a transparent leaderboard that is updated immediately after a user submits their vote. Our platform incorporates a diverse range of evaluation metrics, ensuring that the leaderboard reflects both global model performance and specialized capabilities across various evaluation dimensions. To encourage diverse participation, SWE-Arena allows users to submit a wide range of real-world SE scenarios, without restriction on prompt content. Before accessing our platform, users must agree to the terms of use, including consent for anonymized data collection and public release. This approach balances the need for robust data collection to support research with the protection of user privacy, ensuring that SWE-Arena remains a valuable and ethical resource for the software engineering community.

\section{Future Works}

SWE-Arena represents a significant step forward in the evaluation of FMs for SE tasks. However, there are several opportunities for future enhancements to expand its applicability and improve its effectiveness.

\begin{itemize}
    \item \textbf{Analysis of Real-World SE Workloads}: SWE-Arena plans to analyze user-submitted requests to find common patterns and challenges in software engineering tasks. This analysis could lead to specialized sub-leaderboards focused on particular domains, such as debugging or requirement refinement, and inform the development of more targeted FMs.


    \item \textbf{Enhancing Community Engagement}: SWE-Arena intends to promote contributions from the broader research and development community. This includes enabling users to vote on model performance, contributing to the open-source codebase, and sponsoring initiatives aligned with our platform's objectives.

    \item \textbf{Expanding FM Coverage and Supporting Infrastructure}: SWE-Arena aims to broaden coverage to include domain-specific models and multimodal foundation models, along with infrastructure for more complex SE tasks, including web browsing and API integration.

    \item \textbf{Integrating Advanced Context Compression Techniques}: SWE-Arena intends to incorporate advanced context compression methods to address the challenges of retaining and summarizing long interaction histories. Techniques like LongRope~\cite{dinglongrope} or SelfExtend~\cite{jinllm} could be used to preserve essential context while managing memory constraints, ensuring evaluations accurately reflect real-world SE workflows.
\end{itemize}


\bibliographystyle{plain}
\bibliography{ref}

\end{document}